# Emergence of Self-dual Patterns in Active Colloids with Periodical Feedback to Local Density

Yang Yang[a], Zhi Chao Zhang[a], Fei Qi[b], Tian Hui Zhang[a]

[a]Center for Soft Condensed Matter Physics and Interdisciplinary Research & School of Physical Science and Technology, Soochow University, Suzhou 215006, China.
[b]CAS Key Laboratory of Quantitative Engineering Biology, Shenzhen Institute of Synthetic Biology, Shenzhen Institutes of Advanced Technology, Chinese Academy of Sciences, Shenzhen 518055, China

## Abstract

The central task in the study of self-organization is to explore the general mechanism of emergences. However, this is inhibited by the missing of a full knowledge of the microscopic dynamics of emergence. Here, in this study, the microscopic dynamics of self-organization for patterns is investigated and quantified in a periodically propelled Quincke system. The periodical coupling between propulsion and repulsion at the particle level leads to local directed oscillating particle flows and promises a loop of positive feedback to density fluctuations. Nevertheless, the global evolution of the resulting cluster phase is dominated by a global dual transformation. As stable attractors of the dual transformation, self-dual patterns including stripe patterns and square lattices can be achieved by tuning the strength and the frequency of propelling. However, stripes are possible only at strong propelling where boundary particle flows can form. The findings in this study show that the dynamics of emergence on different length scales are controlled by different mechanisms. The competition and the interplay between different microscopic dynamic processes play the central role in determining the product of emergence. Moreover, the periodically oscillating self-dual patterns demonstrate a classical approach to time crystals.

# Introduction

Complex spatiotemporal patterns, such as spiral galaxy, sand ripples and snowflakes, are ubiquitous in natural and living systems at all scales.[1] Different from equilibrium structures, spatiotemporal patterns generally emerge in systems which are driven out of equilibrium.[2] The studies on self-organization are not only critical in understanding natural phenomena, but also have extensive applications in designing artificial self-organizing systems to fulfill particular functions.[3] To understand and explain self-organizations, various theories of pattern formation have been developed.[4,5] These theories have greatly improved our understanding on the development of embryo, the formation of tissue patterns and the emergence of function for life.[6] However, alternative models concerning different systems with different interactions and dynamics often produce similar patterns.[2,4-6] It follows that mathematical models, which can successfully replicate the patterns observed experimentally, alone cannot clarify what is happening in real experiments. In this case, well-controlled quantitative experiments in which the dynamics of patterning can be monitored are critical. However, pattern formations in natural systems are difficult to control and characterize. To fix the challenges, well-controlled laboratory systems have been developed. Among them, Rayleigh-Bénard convections (RBC),[7] Belousove-Zhabotinsky chemical reactions,[8] Faraday standing-waves[9] and vibrated granular media[10] are the most widely studied examples.

Studies with quantitative experiments have produced a plenty of knowledge of self-organization, and greatly extended our understanding on instability, bifurcation and complexity. However, in spite of its fundamental importance, microscopic dynamics underlying patterning are ubiquitously precluded from observation because of the small length and time scales in atomic systems. In this case, self-organizations in active systems, such as bird flocks,[11] bacteria colonies[12] and self-propelled colloidal systems,[13-16] offer a promising experimental approach for quantitative observations on microscopic dynamics. In these systems, self-organization leads to collective motions such as traveling bands and rotating vortices as an aligning mechanism is present.[14-17] Recent studies[18,19] found that in periodically propelled Quincke systems, dynamic square lattices can emerge from an disordered cluster phase which arising from density fluctuations. These observations demonstrated that active colloids can serve as good experimental models to explore the general principles underlying self-organization with detailed microscopic dynamics. Despite its importance and significance, however, the mechanism which underlies the evolution from the disordered cluster phase to regular patterns was no explored. Most importantly, as many studies have shown that both square lattices and stripes can be achieved in periodically driven systems,[20,21] no stripes were observed in the periodically driven Quincke systems.[19]

Here, we show that both stripe patterns and square lattices are experimentally possible in periodically propelled Quincke systems. As the emergence of square lattices is dominated by the particle flows parallel to density gradient, tangential particle flows at the boundary of a growing cluster, namely boundary flows (BFs), are critical for the emergence of stripes. Moreover, we show that the evolution from disordered cluster phases to regular patterns is dominated by a dual transformation (DT) which results from the combination of well-directed particle flows and well-controlled path-length of individual particles. Driven by the dual transformation (DT), the density distribution becomes dynamically stable till a self-dual pattern is reached. In the steady state, the density distribution oscillates in a half frequency of the propelling between two dual patterns which have the same symmetry and the same structure constants.

## Results

In our experiments (see Methods), dielectric colloidal particles are dispersed in a weakly conducting fluid (Fig. 1a). As a uniform electric field $\vec{E}$ is applied vertically, colloidal particles acquire an antiparallel electric dipole.[22] In the presence of thermal perturbations, a finite in-plane dipole component (perpendicular to the electric field) emerges, and then a net electrostatic torque $\vec{P} \times \vec{E}$ is produced. When the amplitude of the field exceeds a threshold value $E_c$, this torque overcomes the viscous torque and propels the particles to rotate.[23,24] The steady rolling on the bottom surface results in the translational motion at a constant speed $\upsilon$ along a randomly chosen direction. In this study, the Quincke roller are propelled by a square wave field $E$ (SWE) which periodically oscillates between a plateau value $E_p$ ( $>E_c$) and zero (Fig. 1b up). In each cycle, the intervals for $E_p$ and zero are equally set as $0.5T_f$. $T_f$ is the period of the square wave field with a frequency $f$. For Quincke rollers subjected to SWE, they run and stop in the same frequency of field, giving rise the oscillating of speed $\upsilon$ (Fig. 1b bottom) and Brownian behavior (Fig. 1c). At high density, the dipolar interactions between Quincke rollers become significant and can enhance the mobility of paired particles.[13,24] As they are subjected to SWE, the propulsion and the dipolar interaction are switched on and off simultaneously as $E$ oscillates between $E_p$ and 0.

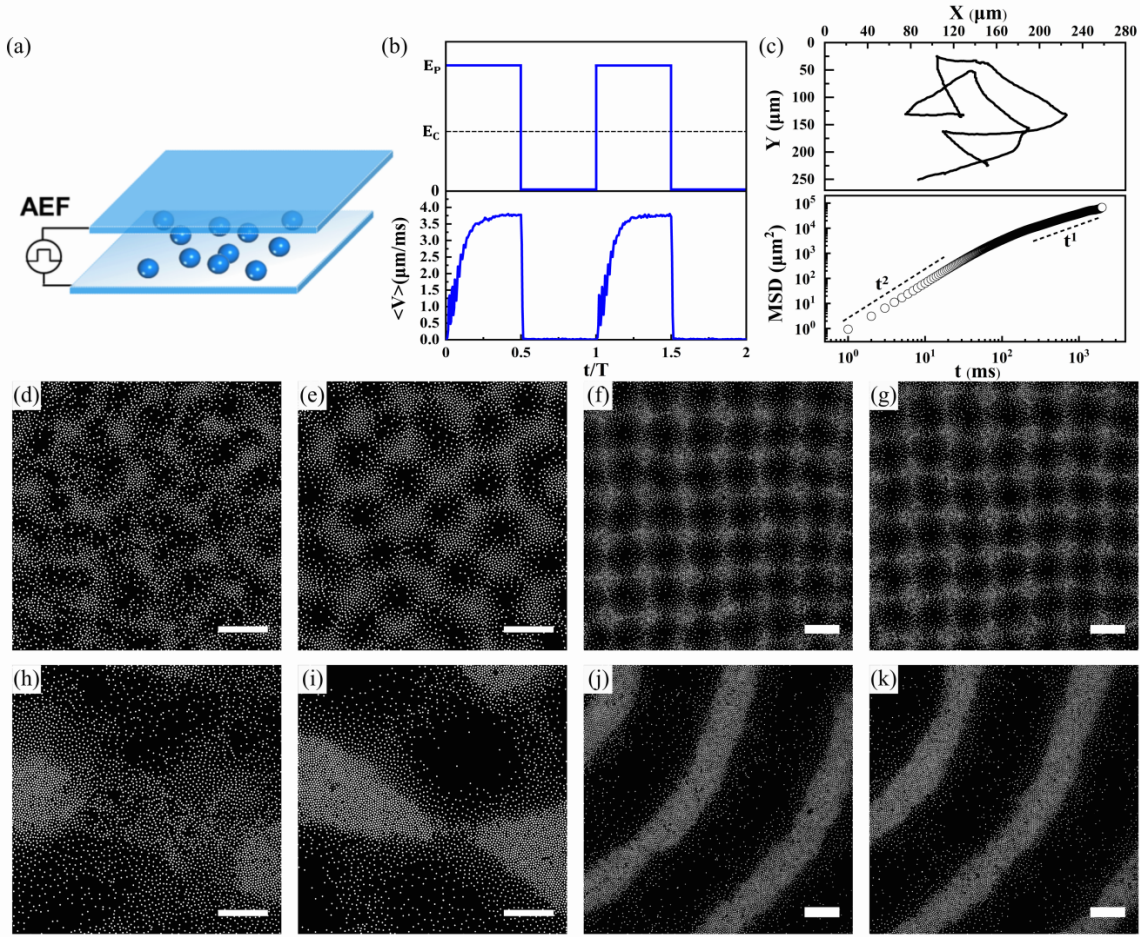

**Figure 1** Emergence of patterns. (a) Experimental setup. (b) Square wave field and the speed of Quincke particles as a function of time. (c) Typical trajectories and the mean squared displacement of Quincke particles subjected to a square electric field of $E=2.1E_c$ and f=5Hz. (d)-(g) Emergence of square lattice at $E=1.8E_c$ and f =30Hz. (d) Density distribution after the first cycle. (c) Density distribution after 20 cycles. (f)-(g) The final steady distributions are represented by a square lattice and its dual partner. (h)-(k) Emergence of stripe pattern at 2.0 $E_c$ and 5Hz. (h) Density distribution after the first cycle. (i) Density distribution after 20 cycles. (j)-(k) Stripe patterns in the final steady state. Scale bar: 100 μm.

Starting from a homogenous state ($\phi_0 = 0.30$), Quincke rollers are set to roll with a randomly selected direction in the first cycle of $E=E_p$. In the running state, density fluctuations emerge quickly. Under a constant field, the coupling between density fluctuations and aligning mechanism leads to the emergence of large-scale collection motions and a dense polar phase.[16,24] However, under a square field, as $E$ drops to zero periodically at the time of $(n-1/2)T_f$, the rollers lose their mobility immediately due to the viscous damping. Here, n=1,2, 3,….., is the number of cycles of SWE. As a result, the density fluctuations developed in the first cycle are frozen at $T_f/2$, giving rise to a non-uniform distribution of clusters (dense domains) and voids (empty domains) (Fig.1d). As $E_p$ is reset in the second cycle, the directions of Quincke rotation in the clusters are not selected randomly anymore because the short-ranged dipolar repulsion between Quincke rollers in the clusters. Driven by the repulsions, rollers are propelled to move from dense regions to spare regions, giving rise to directed particle flows from clusters to voids. The directed particle flows in turn enhance the density fluctuations and leads to denser and larger clusters (Fig.1e). Nevertheless, the amplitude (radius) and the magnitude (internal density) of clusters become saturated after a few cycles (Movie 1). The resulting distribution of clusters is initially disordered (Fig.1e). However, the disordered cluster phase evolves into a dynamic square pattern after hundreds cycles (Fig.1f). The dynamic square pattern oscillates periodically between two square lattices which are dual to each other (Fig. 1f-g) in a frequency half of the SWE (Movie 2). Similar observations haven been reported a recent study.[19] In that study, only square patterns were observed. In our studies, however, stripe patterns can be realized at low frequencies as well. Starting from a homogenous state, the emergence of stripe patterns is preceded by a disordered cluster phase as well (Fig.1h and Movie 3). Nevertheless, the clusters under the conditions for stripe patterns become increasingly elongated (Fig.1i). The elongating clusters eventually become interconnected and form a dynamic superstructure of stripes after hundreds cycle (Fig.1j). In the final steady state, the dynamic pattern oscillates periodically between two stripe patterns which are dual to each (Fig. 1j-k and Movie 4).

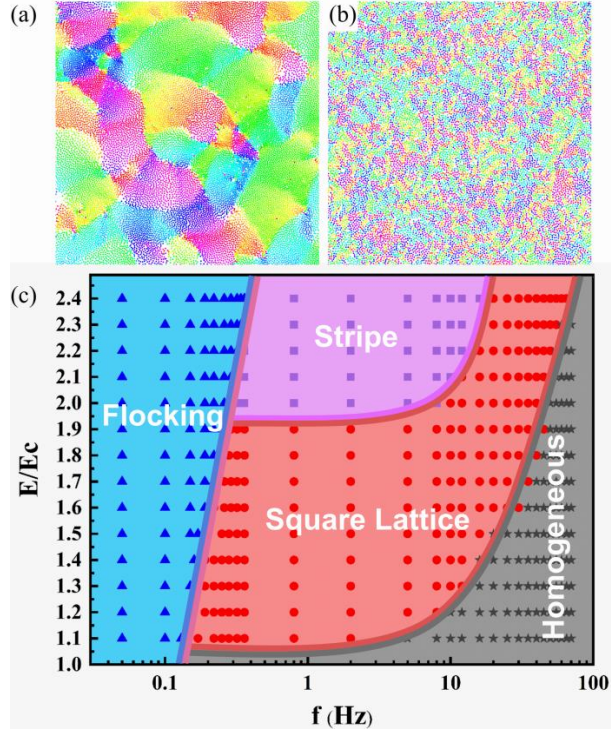

**Figure 2** Phase diagram. (a) Flocking state at $1.9E_c$ and 0.15Hz. $\phi_0$ =0.30. (b) Homogenous state at $1.6E_c$ and 40 Hz. $\phi_0$ =0.30. (c) Phase diagram.

Except the stripe patterns and square lattices, another two states, flocking (Fig. 2a, Movie 5) and homogeneous gases (Fig. 2b, Movie 6), can be realized as well. The phase diagram in $E_p$ - $f$ plane is summarized in Fig. 2c. The square lattices can be achieved at all explored $E_p$. By contrast, the stripe patterns can only be achieved as $E_p$ is above $1.9E_c$ at the global area fraction of $\phi$ =0.3. For $E_p > 1.9E_c$, square lattices are ubiquitously observed at high frequencies while stripe patterns appears only at low frequencies. Given $E_p$, the steady translational speed of Quincke roller is then determined by $\upsilon \sim \sqrt{(E_p/E_c)^2 - 1}$.[16,23] With the well-defined speed, the dynamics of individual Quincke particles at different frequencies are then characterized by different running times and thus different path lengths $S$ traveled by individuals in each cycle. The path length $S$ is defined by $S = \upsilon T_f / 2 = 0.5\upsilon/f$. At high frequencies where $S$ is less than the mean distance *d* between nearest particles, the motion of particles becomes localized and density fluctuations will be suppressed (Movie 5). The trajectories and the mean squared displacement (MSD) in SFig.1 reveal that Quincke rollers in the homogeneous gas states behavior like Brownian particles. At low frequencies, as Quincke rollers have enough time to align their velocity before *E* drops to zero, it is not surprising that they form collective motions (flocking) (Movie 6) as observed in Quincke system subjected to a constant field.[16] The challenge is that in the frequency range between the homogeneous state and the flocking state, what is the mechanism which directs the disordered cluster phases toward a regular pattern and by what mechanism, the value of *S* determines the relative stability of stripes and clusters. To uncover the mechanism underlying the selection of stripe and clusters, it is critical to understand the mechanism which dominates the evolution of clusters phases.

In this study, the instability for the emergence of patterns results from the locally directed particles flows which amplify local density fluctuations. However, directed particle flows alone cannot interpret the evolution from the disordered cluster phases to the regular patterns. We notice that both the stripe patterns and the square lattices are self-dual patterns. In the final steady state, the system oscillates periodically between the patterns and their dual partners via a dual transformation (DT). The symmetry and the structure constants maintain unchanged before and after the DT. In this sense, we suggest that the global disorder-order transition is governed by DT and the final steady patterns represent the stable attractors of DT in phase space. The difference is that the square lattice is a two-dimensional self-dual pattern while the stripe pattern is a one-dimension self-dual pattern. As an important feature of the dual transformation, the particles in each cycle have to be transported from dense domains to void regions, giving rise to the exchange of clusters and voids in space. This is in line with the directed particle flows for the instability. However, to accomplish a dual transformation, it is also critical to control the locations of clusters. In this case, we suggested that the well-defined path length $S$ plays a key role in determining the locations of clusters in DT. The understanding is that driven by $E_p$ and the repulsions from neighbors, particles in clusters start to scatter and move toward the nearest void regions. A well-defined $S$ suggests that the most possible and stable centers for clusters which are going to form in voids are positions which are equally distant from the surrounding existing clusters. Based on this understanding, the centers of clusters in the next cycle can be well predicted with the coordinates of existing clusters. As the first step, the centers of existing clusters are identified by averaging the coordinates of particles included (Fig. 3a, See Materials and Methods). With the existing cluster centers, the plane is divided into Voronoi cells by applying Voronoi tessellation. The edges (blue solid lines in Fig.3a) of Voronoi cells are the bisectors of lines connecting two neighbored cluster centers. The intersection points of edges define the vortices of Voronoi cells. The vortices are equally distant from their source sites (centers of the existing clusters), and therefore the Voronoi vortices represent the most probable positions of clusters in the subsequent cycle. The big red spots in Fig.3b represent the experimentally identified centers of clusters formed in the subsequent cycle. As expected, the vortices are ubiquitously located within the newly formed clusters (Fig.3b). It follows that the evolution of the cluster phase is experimentally dominated by a Voronoi transformation (VT). The structure constructed by the vortices defines the dual partner of the structure constructed by the source sites. In other words, VT in practice defines the DT of a structure. Nevertheless, the centers of new clusters in a disordered distribution do not exactly overlap with the vortices. Therefore, the distribution is not stable under DT. Only in a regular square lattice, the vortices are accurately overlapped with the clusters centers (Fig. 3c-d), and the distribution becomes stable under DT.

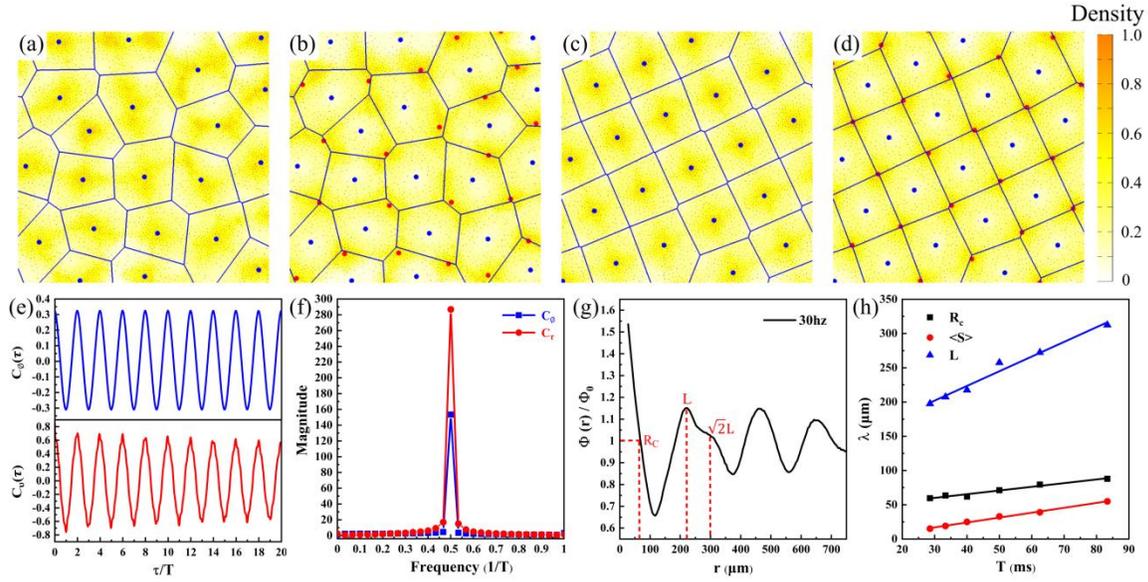

**Figure 3** Dual transformation and spatiotemporal properties of square lattices at $1.8E_c$ and 30Hz. $\phi_0 =0.30$. (a)-(b): Voronoi diagram of the cluster phase. (c)-(d): Voronoi diagram in the steady square lattice. Big blue spots: the center of clusters. Small black spots: Static particles. Big red spots: Centers of clusters formed in the subsequent cycle. (e): Density time correlation and velocity time correlation in the oscillating square lattices. (f): Frequency spectrum of the time correlations. (g): Radial density distribution in the square lattices. (h) Dependence of lattice constants on the period of square field.

In a steady square lattice, the density distribution oscillates between the two dual lattices such that the local density oscillates between a cluster state and a void state. As a result, the autocorrelation $C_\phi(\tau)$ (see Methods) of local density also exhibit oscillatory behavior (Fig.3e up). Accompanying DT, particles go and return periodically between neighbored clusters and voids. As a reflection of the periodic motion, the velocity autocorrelation exhibits an oscillatory behavior as well (Fig. 3e bottom). The frequency spectrums of the autocorrelations show that both the density and the velocity are oscillating with a half frequency (or a doubled period) of the field (Fig. 3f). In periodically driven quantum systems, the emergence of a half frequency oscillation in spin offers an experimental approach for time-crystal.[25,26] However, the exploration of time crystals so far is mainly focused on Quantum systems and Quantum effects play a critical role. It is still a challenge to realize time crystals in classical systems. Here, the oscillating square lattices demonstrate a classical time-crystal in terms of the local density.

To quantify the characteristic sizes of the square lattice, the radial distribution of density (area fraction) around the centers of clusters $\phi(r)$ is calculate and averaged over all centers. The resulting $\langle \phi(r) \rangle$ is then is normalized by the global value $\phi_0$. Starting from the cluster center, the density decreases with r (Fig. 3g). The radius, $R_c$, where $\langle \phi \rangle / \phi_0 = 1$ is identified as the characteristic radius of the clusters. Beyond $R_c$, the density oscillates and exhibits a series of peaks. The first peak consists of two sub-peaks. The first sub-peak represents the mean center-center distance between two nearest clusters and defines the lattice constant $L$. The second sub-peak arises from the next nearest clusters. The location of the second sub-peak, $\sqrt{2}L$, suggests that the lattice has a square symmetry. As the period $T_f$ increases upon the decrease of frequency, $S$ increases linearly with $T_f$ as expected. Both $R_c$ and $L$ increase linearly with $T_f$ as well. Nevertheless, the linear dependence of $R_c$ and $L$ on $T_f$ is interrupted as frequency goes down to the stripe regions. Both square lattices and stripe patterns can be produced by DT. Therefore, the mechanism of DT alone cannot account for the emergence of stripes. There should be an additional mechanism which does not work at the conditions of square lattices but play a role in the emergence of stripes. We notice that the emergence of stripes is preceded by a cluster phase as well. Distinct from that in the evolution for square lattices, the clusters become elongated increasingly. The mechanism which underlies the elongation is the central to understand the transition from cluster to stripe.

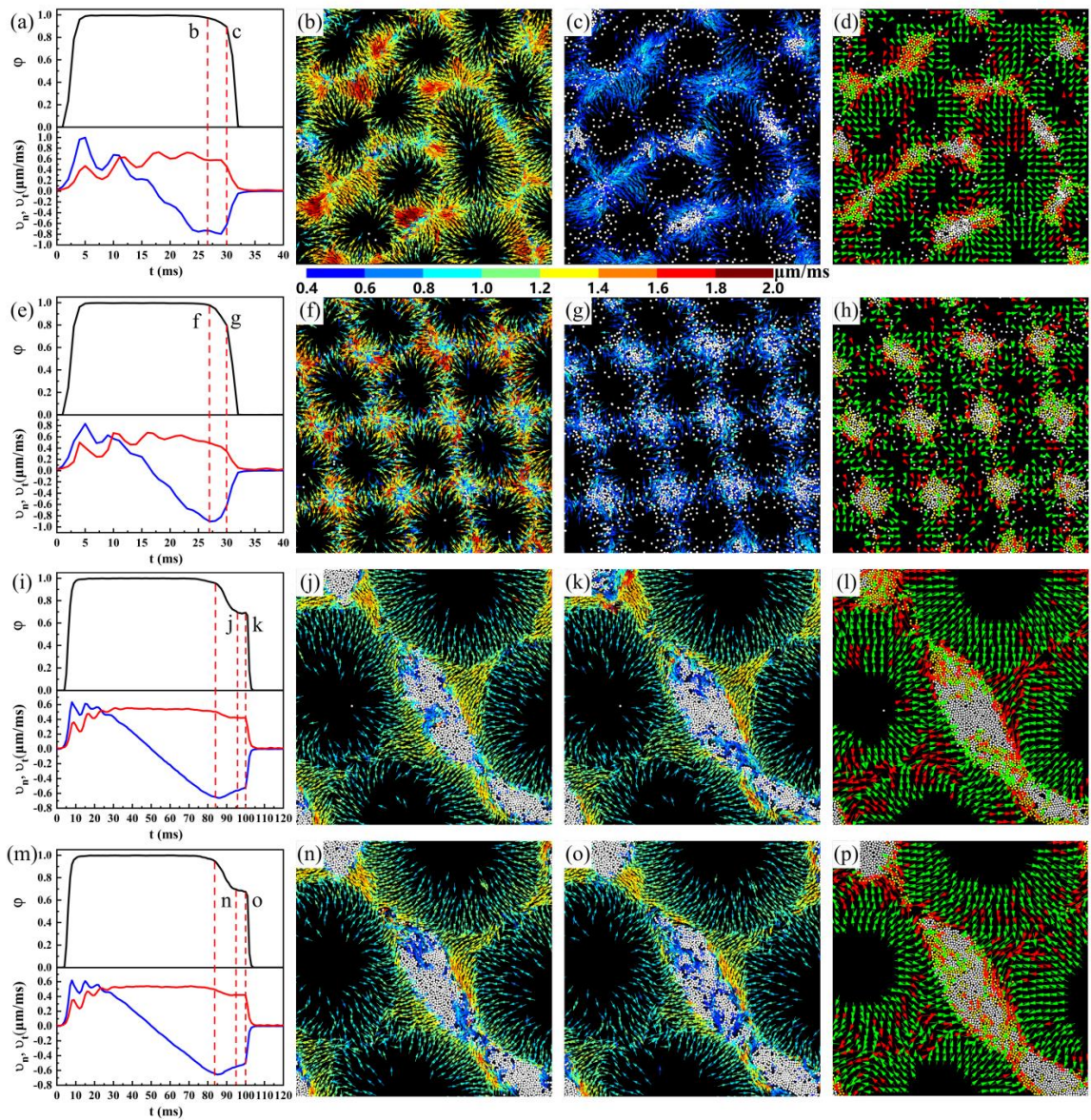

**Figure 4** Particle flows in clustering processes at 2.1$E_c$ and $\phi_0$ =0.30. (a)-(h): $f$=16Hz. (i)-(p): $f$=5Hz. (a) Fraction of active particles and the mean magnitude of particle flows in the disordered cluster phase in the emergence of square lattice. (b)-(c): Particle flows around a growing cluster. (d): The corresponding flow field of (c). Green arrow: $\vec{v}_n > \vec{v}_t$ arrow. Red arrow: $\vec{v}_n < \vec{v}_t$. White spots: static particles. Yellow spots: Mobile particles in clusters. (e)-(h): Particle flows in the steady square lattice. (i)-(l): Particle flows in the disordered cluster phase in the emergence of stripes. (m)-(p): Particle flows after 2$T_f$.

The elongation of clusters requires redistribution of particles. Therefore, a full information of particle flows during the growth of clusters is critical to understand the mechanism of elongation. To quantify the particle flows, local density gradient $\nabla \phi(\vec{r})$ is calculated. The unit vector $\vec{n} = -\nabla \phi(\vec{r}) / |\nabla \phi(\vec{r})|$ defines the local direction pointing from high density to low density. The quality $\vec{v}_n = \vec{v} \cdot \vec{n}$ defines the component of velocity along $\vec{n}$ and offers a measurement on local directed particle flows. The quality $\vec{v}_t = \vec{v} - \vec{v}_n$ defines the component of particle flows perpendicular to the density gradient. Starting from the homogeneous states at $E_p = 2.1 E_c$, the dynamic evolutions of clusters at f=16Hz (square lattice) and 5Hz (stripe) are investigated respectively with quantified $\vec{v}_n$ and $\vec{v}_t$. At the beginning of $E_p$, particles move from clusters to voids, giving rise to positive $\vec{v}_n$. As particles begin to aggregate in the voids, $\vec{v}_n$ becomes negative. Figure 4a represents the results observed in the cluster phase evolving toward square lattices. The quantity $\varphi$ defines the fraction of active particles and the values of $\bar{v}_n = <\vec{v}_n>$ and $\bar{v}_t = <\vec{v}_t>$ give the mean magnitude of the normal and tangential particle flows. The fraction $\varphi$ of active particles increases sharply from zero to 1 as $E_p$ starts and drops to zero at the end of $E_p$. Nevertheless, the decrease of $\varphi$ consists of two distinct linear regions. The corresponding snapshots (Fig.4b-c) show that the initial decrease of $\varphi$ is induced by clustering: The particles begin to aggregate and form clusters in the void regions. The growth of clusters leads to the increase of static particles and thus the decrease of $\varphi$. The linear decrease (from state b to state c) continues for around 7ms and then suddenly drops to zero at the ending of $E_p$ when all particles lose their mobility quickly due to damping (Movie 7). Accompanying the decrease of $\varphi$, $\bar{v}_n$ and $\bar{v}_t$ start to decrease simultaneously (Fig.4a bottom). Similar scenario is observed in the steady lattice (Fig.4e-g and Movie 8). As a common feature, the particle flows in the clustering processes are dominated by the normal component $\bar{v}_n$ ($> \bar{v}_t$) although tangential component $\bar{v}_t$ is also significant. However, in the disordered cluster phase where both the shape and the distribution of clusters are irregular, $\bar{v}_t$ can be locally stronger than $\bar{v}_n$ (Fig.4c-d). It is found that the tangential component $\bar{v}_n$ plays a key role in shaping clusters and adjusting their spatial arrangement.

In the cluster phase evolving toward stripes, the clustering processes leads to the decrease of $\varphi$, $\bar{v}_n$ and $\bar{v}_t$ (Fig.4i) as well. However, the magnitudes of $\bar{v}_n$ and $\bar{v}_t$ becomes comparable in clustering. Most interestingly, the decrease of $\varphi$ in the clustering processes is intermediated by a steady state where the fraction of active particles maintains unchanged till the ending of $E_p$. It suggests that after the initial clustering, the active particles arriving at the boundaries of growing clusters don't lose their mobility anymore but keep moving. Figure 4j-k illustrate that corresponding to the flat state, local tangential particle flows are established at the boundaries of clusters after the initial clustering such that the incoming particles begin to flow along the boundaries of clusters. The emergence of boundary flows (BFs) results in the steady $\varphi$. As a reflection of BFs, the local flow fields are dominated by tangential component $\bar{v}_t$ (Fig.4l). The direction and the magnitude of BFs are generally determined by local dynamics such that they are sensitive to local fluctuations and are generally not uniform in all directions around a cluster (Movie 9). As a growing cluster is subjected non-uniform BFs, particles are transported along the boundaries asymmetrically such that the shape of clusters become unstable and get elongated in the dominant directions of BFs. As a result, the clusters become elongated increasingly and get interconnected as Fig.4n-p illustrate. Moreover, SFig.2a-d shows that in the steady oscillating stripe patterns, BFs are still significant and play a key role in stabilizing the dynamic stripes (Movie 10).

The results in Fig.4 demonstrate that the emergence of BFs is the key for the transition from cluster to stripe. In the presence of BFs, clusters are not stable with respect to stripes. However, before the formation of BFs, a static cluster has to be form first. After the formation static clusters, it is also critical that the particle flows around them last for a finite time to be redirected. The understanding is that due to the interaction between the growing clusters and the incoming particles, the hydrodynamic flows will be blocked and reflected at the boundary of clusters. The interaction between the reflected hydrodynamic flows and incoming flows will modify and redirect the local flows. However, a finite responding time $\tau_f$ is necessary to rebuild a steady flow field. The interval for the linear decrease of $\varphi$ before the steady state offers an experimental estimation for $\tau_f$. At $E_p$=2.1$E_c$, $\tau_f$ is around 10ms (Fig.4i and Fig. 4m). Upon the decrease of frequency, clusters become larger and thus the clustering time $\tau_c$ As $\tau_c > \tau_f$, the clusters become unstable due to the emergence of BFs and only a stripe pattern can persist. In the cluster phase of square lattice, the clustering process is too short to form BFs (Movie 7 and Movie 8). Experimentally, however, stripe patterns can only form and persist within a finite range of frequencies. As the frequency is below a critical value, stripes are not stable with respect to flocking.

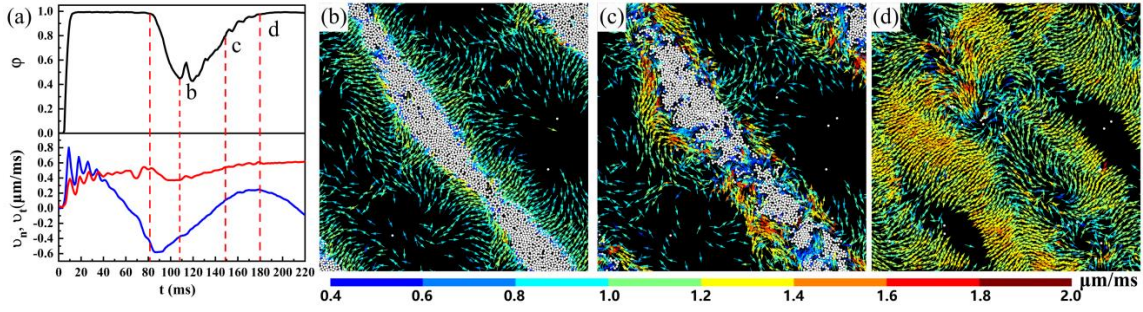

**Figure 5** Particle flows in the transition from stripe to flocking at $2.1E_c$ and $\phi_0$ =0.30 from 5Hz to 0.15Hz.

To uncover the dynamics underlying the transition from stripe to flocking, a stable stripe pattern achieved at $2.1E_c$ and 5Hz is subjected to a SWE of 0.1Hz at the same $E_p$. It shows that stripes form as well at the time of 100ms (Fig. 5b), which is the point where $E_p$ stops in the case of 5Hz. In the case of 0.15Hz, however, particles are continuously transported by BFs along the boundaries after the time of 100ms. Simultaneously, it is found that more and more static particles in the stripes are reactivated to move (Fig.5c). As stripes melt completely (Fig.5d), flocking begin to emerge (Movie 11). It follows that the melting of stripes arises from the reactivation of particles which have been stopped by colliding and damping. To find out the time $\tau_a$ for reactivation, the behavior of $\varphi$ in the melting process is investigated. The moment where $\varphi$ begins to decrease states the birth of stripes and the time when $\varphi$ goes back to 1.0 indicates the complete melting. The interval between the birth and the melting serves an measurement for the maximum lifetime or the reactivation time $\tau_a$ of static clusters. Theoretically, $\tau_a$ should be a function of $E_p$. Experimentally, however, we suggest that the size and the internal density of stripes may also affect the value of $\tau_a$.

Based on the observations, we conclude that the competition between dynamic processes of clustering, emergence of BFs and reactivation plays the central role in determining the result of emergence in a periodically propelled Quincke system. The maximum lifetime $\tau_a$ of a static cluster defines the longest time permitted for clustering. As the clustering time $\tau_c$ is shorter than $\tau_a$, regular patterns are possible. As the clustering time $\tau_c$ is less than the time $\tau_f$ for BFs, square lattices are stable. As $\tau_c > \tau_f$, BFs emerge and stripes become more stable with respect to clusters. However, to achieve a stable stripe pattern, it is also necessary that $\tau_f < \tau_a$. Otherwise, all stopped particles will be reactivated such that no pattern is possible.

    The phase diagram in Fig. 2 shows that as $E_p < 1.9E_c$, no stripe phase exists. We suggest that in a viscous solvent, because of the dissipation, low-mobility particles will lose most or all their kinetic energy during the colliding at the boundaries of clusters. In this case, it is difficult or much more time is necessary to form dynamic BFs. For Quincke rollers, the mobility in terms of speed is determined by $E_p$. Therefore, stripe patterns are difficult at low field and square lattices are the only possible patterns in experiments. This also interprets the observation in a recent study[19] where no stripe patterns had been realized as $E_p$ was fixed at a value which we believe is below the critical field for BFs. To demonstrate the role of $E$ in the formation of BFs. Observations is conducted at 2Hz and 1.3$E_c$. As $E_p$ is higher than 1.9$E_c$, stripe patterns will form at 2Hz. At $E=1.3E_c$, however, a square lattice of large clusters form. As expected, the fraction of active clusters in clustering (Fig. 6a) decreases continuously and exhibits no a steady state before the ending of $E_p$ although the clustering process persists for more than 50ms. This can be seen directly in the snapshots of velocity field (Fig.6b-d, Movie 12). It also demonstrates that the typical time for BFs is dependent on the field strength.

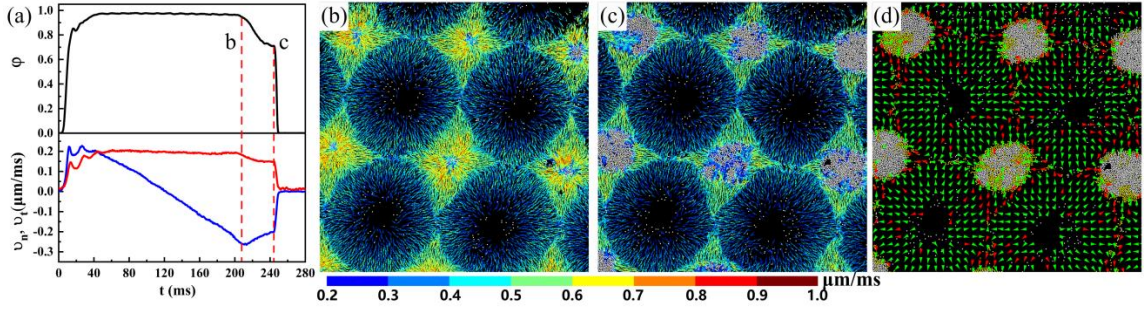

**Figure 6** Particle flows in the square lattice 1.3$E_c$ and 2Hz. $\phi_0$ =0.30.

Both stripe patterns and square lattices have been observed in vertically vibrating liquid films[20] and granular systems[21]. In these systems, the patterns are generally understood as steady solutions of standing waves which can survive with the given boundary conditions.[2,5] However, the mechanisms underlying the global disorder-order transition are unclear so far due to the challenge of experimental observations on microscopic dynamics. The findings here that BFs can destabilize a cluster phase and trigger the transition from square lattice to stripe pattern may be applied to understand the emergence of stripes in vibrated granular systems.

## Conclusions

In summary, the coupling between propulsion and repulsion in a periodically propelled Quincke system results in directed particle flows parallel to local density gradient, giving rise to the instability of a uniform state. The magnitude and the amplitude of density fluctuations are limited by the path length travelled by particles in one cycle. The saturated density fluctuations result in disordered cluster phases. Driven by DT, the topological properties of cluster phases become stable till self-dual patterns emerge as the stable attractors of DT. The oscillating stripe patterns and square lattices offer two classical examples of time crystal in which dense domains melt and form periodically. In the clustering process, the competing and the interplay between damping, redirecting and reactivation of Quincke rollers determines the final product of DT. As the clustering time $\tau_c$ is larger than the redirecting time $\tau_f$ for BFs, cluster phases are not stable with respect to stripes due to the emergence of BFs. When the clustering time is even larger than the reactivation time $\tau_a$, no stable patterns can form and flocking emerges. The time scales of the microscopic dynamic processes are strongly dependent on the field strength $E_p$ of SW. Here, the formation of cluster and the evolution from disorder cluster phases to regular patterns are dominated by local directed particle flows and a global DT respectively. It follows that the dynamics of emergence at different length scales are controlled by different mechanisms. However, the global DT is still closely related to microscopic properties: well-directed motion and well-controlled traveling length of individual particles. Therefore, a full understanding of microscopic dynamics is critical in controlling and predicting the results of emergence. In this case, well-controlled active colloids offer a robust experimental model system which permits quantified observation at single-particle level.

## Materials and Methods

**Experimental Setup.** Colloidal particles (Thermo Scientific G0500, polystyrene spheres) with diameter 4.90$\mu$m are dispersed in the mixture of 0.15M AOT/hexadecane. The suspension is sealed in a cell constructed by two ITO-coated glass slides which are separated by insulating glass spacer (Fig. 1a). The square wave field is amplified by a voltage amplifier. The dynamic processes are observed with a Nikon microscope and recorded by a SCMOS camera at a rate of 500 (or 1000 frames) per second. Using IDL software, the positions of particles are located and tracked with an accuracy of 0.1 pixel.[27]

**Cluster centers.** In the cluster phase, particles which have a local area fraction larger than the global mean value $\phi_0$ are identified as ones belonging to a cluster. The cluster centers are then identified by averaging the coordinates of all particles include.

**Velocity autocorrelation**

With the velocity $\vec{v}_i(t)$ of particles, the velocity autocorrelation is calculated and averaged over all time and particles.

$$C_v(\tau) = \left\langle \frac{1}{N}\sum_{i=1}^{N} \vec{n}_i(t)\cdot\vec{n}_i(t+\tau) \right\rangle_t \text{ and } \vec{n}_i(t) = \frac{\vec{v}_i(t)}{|\vec{v}_i(t)|}$$

$N$ is the number of particles.

**Density time correlation**

$$C_\phi(\tau) = \left\langle \frac{1}{M}\sum_{i=1}^{M} \Delta\phi(\vec{r}_i,t)\Delta\phi(\vec{r}_i,t+\tau) \right\rangle_t \text{ and } \Delta\phi(r,t) = \left(\phi(\vec{r},t)-\phi_0\right)/\phi_0$$

Here, $\phi_0$ is the global mean area fraction of the system. The system is gridded with the radius $R_c$ identified in Fig. 3g. M is the number of grids.

## Author Contributions

T.H.Z. designed research; Z.C.Z. and Y.Y. performed experiments; Y.Y., F.Q. and T.H.Z. analyzed data; and T.H.Z. wrote the paper.

## Conflicts of interest

There are no conflicts to declare.

## Acknowledgement

We thank Lei Han Tang, Hugues Chate and Xia-qing Shi for interesting discussions, and for a critical reading of the manuscript and useful suggestions. T.H.Z. acknowledges financial support of the National Natural Science Foundation of China (Grant 11974255 and 11635002.).

# Supplementary Materials

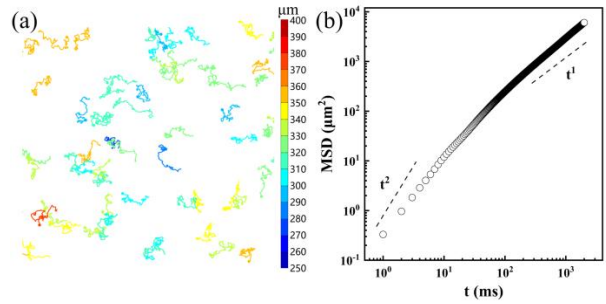

**SFigure 1** Dynamic behavior of Quincke particles in the homogeneous state) at $1.6E_c$ and 40 Hz. $\phi_0$ =0.30

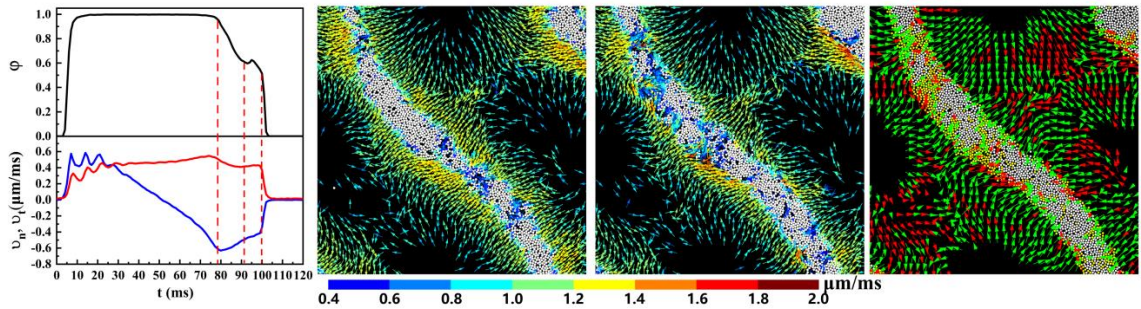

**SFigure 2** Particle flows in the steady stripe patterns at $2.1E_c$ and 5Hz $\phi_0$ =0.30.

# Movie description

**Movie 1**: Cluster phase of square lattice at 1.8Ec and 30Hz. Area fraction $\phi$=0.30.

**Movie 2:** Square lattice at 1.8Ec and 30Hz. Area fraction $\phi$=0.30.

**Movie 3:** Cluster phase of stripe pattern at 2.0Ec and 5Hz. Area fraction $\phi$=0.30.

**Movie 4:** Stripe pattern at 2Ec and 5Hz. Area fraction $\phi$=0.30.

**Movie 5:** Flocking at 2.1Ec and 0.15Hz. Area fraction $\phi$=0.30.

**Movie 6:** Homogeneous gas at 1.6Ec and 40Hz. Area fraction $\phi$=0.30.

**Movie 7:** Particle flows in the disordered cluster phase at 2.1Ec and 16Hz. Area fraction $\phi$=0.30. Running particles are represented by arrows and the static particles in clusters are represented by white dots.

**Movie 8:** Particle flows in the steady square lattice at 2.1Ec and 16Hz. Area fraction $\phi$=0.30. Running particles are represented by arrows and the static particles in clusters are represented by white dots.

**Movie 9:** Particle flows in the disordered cluster phase at 2.1Ec and 5Hz. Area fraction $\phi$=0.30. Running particles are represented by arrows and the static particles in clusters are represented by white dots.

**Movie 10:** Particle flows in the steady stripe pattern at 2.1Ec and 5Hz. Area fraction $\phi$=0.30. (colored velocity). Running particles are represented by arrows and the static particles in clusters are represented by white dots.

**Movie 11:** Particle flows in the transition from stripes to flocking at 2.1Hz.article flows in the disordered cluster phase at 2.1Ec and 0.15Hz. Area fraction $\phi$=0.30. Running particles are represented by arrows and the static particles in clusters are represented by white dots.

**Movie 12:** Particle flows in the steady square lattice at 1.3Ec and 2Hz. Area fraction $\phi$=0.30. Running particles are represented by arrows and the static particles in clusters are represented by white dots.